\documentclass[epj]{svjour}

\usepackage{amsmath,amssymb,latexsym,bm}
\usepackage{citesort}
\usepackage{graphics}

\newcommand{\dif}{{\rm d}}

\newcommand{\abar}{\bar{\alpha}_s}
\newcommand{\atpi}{\frac{\abar}{2\pi}}
\newcommand{\rpa}{\mathcal{R}_{\rm pA}}
\newcommand{\mean}[1]{\left\langle #1 \right\rangle_Y}
\newcommand{\dk}[3]{\frac{(\bm{#1}-\bm{#2})^2}
    {(\bm{#1}-\bm{#3})^2 (\bm{#3}-\bm{#2})^2}}

\begin{document}

\title{From Classical to Quantum Saturation
    in the Nuclear Gluon Distribution\thanks{Based on talk given at
    {\sl ``Hard And Electromagnetic Probes Of
    High Energy Nuclear Collisions''}, Ericeira,
    Portugal, November 2004}}

\author{D.N.~Triantafyllopoulos\inst{}\thanks{{\it E-mail address:}
dionysis@dsm-mail.saclay.cea.fr}}

\institute{SPhT/Saclay/CEA, 91191 Gif sur Yvette, France}

\date{14 February 2005}

\abstract{We study the gluon content of a large nucleus (i) in the
semi-classical McLerran-Venugopalan model and (ii) in the high
energy limit as given by the quantum evolution of the Color Glass
Condensate. We give a simple and qualitative description of the
Cronin effect and high-$p_{\rm \scriptscriptstyle{T}}$ suppression
in proton-nucleus collisions.
    \PACS{
      {11.10.Hi}{} \and
      {12.38.-t}{} \and
      {13.85.Lg}{} \and
      {24.85.+p}{}     } }

\maketitle

\section{Introduction}
\label{intro} At high energies and/or for large atomic number $A$,
the wavefunction of a hadron is expected to be dominated by a high
density gluonic system. Gluons having occupation numbers $\varphi$
of order $1/\alpha_s$, which is the maximal density allowed by their
mutual interactions, overlap in phase space and saturate
\cite{GLR,MV}. A strong classical field is associated with the
wavefunction and assumes a value $\mathcal{A} \sim \sqrt{a^{\dagger}
a} \sim \sqrt{\varphi} \sim 1/g$ at saturation. At the same time
scattering amplitudes become of order 1 and unitarity limits are
reached \cite{Mue95}. The problem can be attacked by weak coupling
methods, since the non-linear phenomena ``push'' gluons to occupy
higher momenta, and the saturation momentum $Q_s$, which is defined
as the scale where $\varphi(Q_s) \sim 1/\alpha_s$, is a hard scale
increasing as a power of energy in the small Bjorken-$x$ limit.

Presumably one of the most complete and modern approaches to
saturation is the effective theory of the Color Glass Condensate
(CGC) \cite{ILM01ab,FILM02}. Fast moving partons have a large
lifetime due to time dilation and act as ``frozen'' sources $\rho$
for the virtual emission of softer gluons. One solves the classical
Yang-Mills equations to obtain the color field $\mathcal{A}(\rho)$
and then an observable $O(\mathcal{A})$ is determined by averaging
over the possible color sources, with a probability distribution
$W_Y[\rho]$. Increasing the rapidity $Y=\ln(1/x)$, more gluons need
to be included in the source, and a resummation of $\alpha_s Y$
enhanced terms in the presence of a background field leads to a
functional Renormalization Group Equation (RGE) for $W_Y[\rho]$
\cite{ILM01ab,FILM02,Wei02,JKLW97}. This RGE gives an infinite
hierarchy of non-linear coupled equations, the so-called Balitsky
equations \cite{Bal}. The first one describes the evolution of the
scattering amplitude $\mean{T_{\bm{x}\bm{y}}}$ of a color dipole
$(\bm{x},\bm{y})$ off the CGC and reads
\begin{align}\label{Bal1}
    \frac{\partial \mean{T_{\bm{x}\bm{y}}}}
    {\partial Y}\,=
    \atpi &\int \dif^2 \bm{z}\,
    \dk{x}{y}{z}
    \nonumber \\
    &\times \mean{ T_{\bm{x}\bm{z}} + T_{\bm{z}\bm{y}}
    - T_{\bm{x}\bm{y}} - T_{\bm{x}\bm{z}} T_{\bm{z}\bm{y}}},
\end{align}
with $\abar\!=\!\alpha_s N_c/\pi$. The first three terms correspond
to the BFKL equation \cite{BFKL} in coordinate space \cite{Mue94},
while the last one accounts for unitarization effects.
Eq.~(\ref{Bal1}) can be closed by a mean field approximation, that
is by allowing the last term to factorize\footnote{However, this
factorization is not valid in the region where the amplitude is very
small, namely when $T \lesssim \alpha_s^2$
\cite{MS04,IMM05,IT0405,MSW05}.}, something which should be
reasonable assuming that the target is a large ($A \gg 1$) nucleus
\cite{Kov99}.

\section{Classical Saturation}
\label{classical} Classical saturation, where there is no small-$x$
evolution, can be realized only in a large nucleus. The $A \times
N_c$ valence quarks are the sources for the emission of gluons, and
in the McLerran-Venugopalan (MV) model \cite{MV} they are assumed to
be uncorrelated for transverse separations $\Delta \bm{x} \lesssim
\Lambda_{\rm QCD}^{-1}$, so that the probability distribution is
given by the Gaussian \cite{MV,Kov96a}
\begin{equation}\label{WMV}
    W_{\rm MV}[\rho] \propto \exp
    \left[
    -\frac{1}{2} \int^{1/\Lambda} \dif^2 \bm{x}\,
    \frac{\rho_a(\bm{x})\rho_a(\bm{x})}{\mu_A^2}
    \right],
\end{equation}
where $\mu_A^2 = 2 \alpha_s A / R_A^2 \sim A^{1/3} \Lambda^2$ is the
color charge density squared, with $R_A$ the nuclear radius. Even
though the sources are uncorrelated, the created field $\mathcal{A}$
is obtained from a non-linear equation. Thus, starting from its
canonical definition, the gluon occupation number $\varphi_A$ is not
the one that we would obtain from a simple superposition of the
sources. When $A \gg 1$, the density can be high, sets the magnitude
of the saturation scale as $Q_s^2(A) \approx \Lambda^2 A^{1/3} \ln A
\gg \Lambda^2$, and we find \cite{IIT04}
\begin{equation}\label{phi}
    \varphi_A =
    \frac{1}{\alpha_s}\,
    {\rm \Gamma}(0,z) +
    \varphi_A^{\rm twist}(z), \qquad z
    \equiv k^2/Q_s^2(A),
\end{equation}
with $\bm{k}$ the transverse gluon momentum and $k$ its magnitude.
Here, ${\rm \Gamma}$ is the incomplete Gamma function, while the
explicit expression for $\varphi_A^{\rm twist}$ can be found in
\cite{IIT04}.

\begin{figure}[t]
    \centerline{\resizebox{0.5\textwidth}{!}
    {\includegraphics{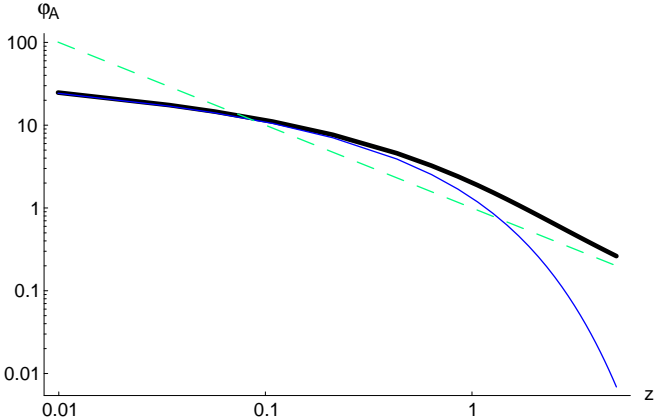}}}
    \caption{\sl
    The gluon occupation number in the MV model. Thick (black),
    solid (blue) and dashed (green) lines show the total,
    CGC and BS quantities respectively.}
    \label{fig1}
\end{figure}

\begin{figure}[t]
    \centerline{\resizebox{0.5\textwidth}{!}
    {\includegraphics{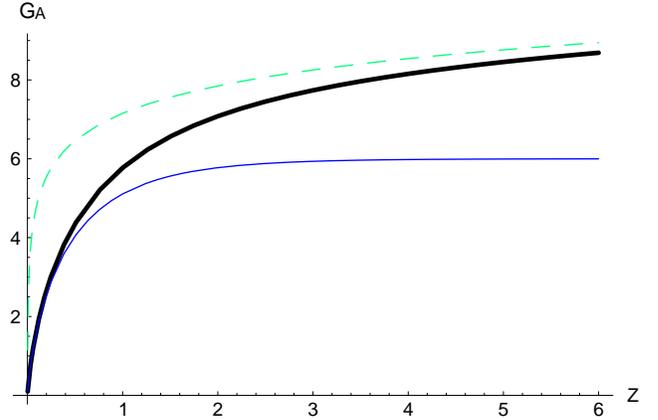}}}
    \caption{\sl
    The integrated gluon distribution in the MV model.
    Thick (black), solid (blue) and dashed (green) lines show the
    total, CGC and BS quantities respectively.}
    \label{fig2}
\end{figure}

The first term, enhanced by $1/\alpha_s$, dominates for all $z
\lesssim 1$, as shown in Fig.~\ref{fig1}. We interpret this compact
distribution, falling exponentially at large $z$, as the occupation
number in the CGC phase. The twist term\footnote{The coefficient in
front of this term is $\{\alpha_s \ln [Q_s^2(A)/\Lambda^2]\}^{-1}$,
which is assumed to be equal to one. In fact, in a running coupling
treatment of the problem, this identification becomes natural
\cite{IIT04}.} contains the bremsstrahlung spectrum (BS)
$\varphi_{\rm BS} \sim 1/z$, and is important for the large-$z$
behavior, while it remains finite as $z\rightarrow 0$. Due to the
lack of correlations among the valence quarks, a sum rule exists
\cite{MV,GP02,IIT04,KKT03}
\begin{equation}\label{sumrule}
    \int^{Z} \dif z \,\left[\varphi_A(z)-\varphi_{\rm BS}(z)\right]
    \xrightarrow[Z \rightarrow \infty]{} 0;
\end{equation}
the integrated distribution is obtained by ``summing'' over the
nucleons when $Q^2 \equiv Z Q_s^2(A) \gg Q_s^2(A)$ (see
Fig.~\ref{fig2}). Thus, the effect of the repulsive interactions in
the nucleus is just a redistribution of the gluons in momenta. The
spectra $\varphi_A$ and $\varphi_{\rm BS}$ become equal at a scale
$Q_c(A)$ such that $\Lambda^2 \ll Q_c^2(A) \approx \alpha_s Q_s^2(A)
\ll Q_s^2(A)$ and ``infrared'' gluons in excess in the BS spectrum
are located at $k \sim Q_s(A)$ in the MV model. Therefore, \emph{the
MV spectrum is enhanced around the saturation scale}, as shown in
Fig.~\ref{fig1}.

\begin{figure}[b]
    \centerline{\resizebox{0.5\textwidth}{!}
    {\includegraphics{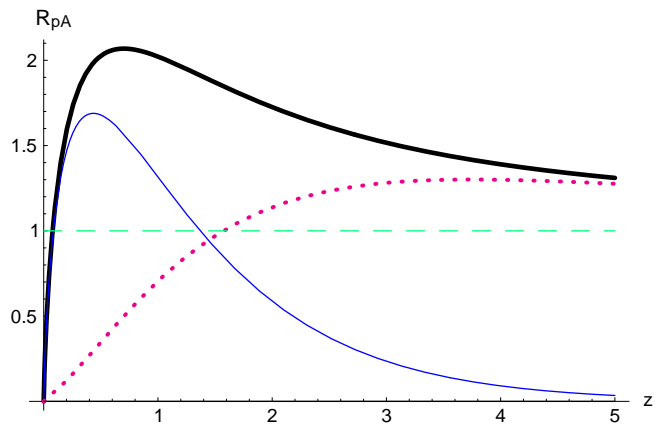}}}
    \caption{\sl
    The Cronin ratio in the MV model. Thick (black), solid (blue)
    and dotted (red) lines show the total, CGC and twist
    contributions respectively.}
    \label{fig3}
\end{figure}

As an immediate consequence, let us consider the Cronin ratio
\begin{align}\label{RpA}
    \rpa \equiv
    \frac{\varphi_A}{A^{1/3}\varphi_p} =
    \frac{\varphi_A}{\varphi_{\rm BS}} = z\, \varphi_A,
\end{align}
with $\varphi_p$ the spectrum of the proton, when this is obtained
from a simple superposition of the gluons emitted by its valence
quarks. It behaves as
\begin{align}
    &\bullet \rpa \ll 1 \hspace{-1.25cm}&
    \rm{if} &\;\; z \ll 1 \nonumber \\
    &\bullet \rpa \sim \mathcal{O}(1/\alpha_s) \gg 1
    \hspace{-1.25cm}&
    \rm{if} &\;\; z \sim 1\\
    &\bullet \rpa \rightarrow 1^+ \hspace{-1.25cm}&
    \rm{if} &\;\; z \gg 1. \nonumber
\end{align}
The ratio, shown in Fig.~\ref{fig3}, has a maximum at
$z_m\hspace{-0.1cm}=\hspace{-0.1cm}0.435 + \mathcal{O}(\alpha_s)$
\cite{IIT04}. The maximal value $\rpa^{\rm max}=0.281/\alpha_s +
\mathcal{O}({\rm const})$ corresponds to a pronounced peak
\cite{IIT04,GP02,GJ03} originating from the compact nature of the
nuclear wavefunction at saturation. This value increases with $A$
(since $1/\alpha_s \equiv \ln Q_s^2(A)/\Lambda^2$).

\section{Quantum Saturation}
\label{quantum}

Now consider the evolution of a hadron to higher energies. Its
wavefunction contains more and more soft gluons, due to the
$\alpha_s Y$ increase in the available longitudinal phase space, and
correlations among the color sources are induced. The gluon
occupation number may be obtained from the (averaged over impact
parameter) dipole-hadron scattering amplitude as
\begin{align}\label{phiT}
    \varphi(\bm{k},Y) =
    \frac{1}{\alpha_s} \int
    \frac{\dif^2 \bm{r}}{\pi \bm{r}^2}\,
    \exp(-i \bm{k}\! \cdot \!\bm{r})\, T(\bm{r},Y),
\end{align}
where $\bm{r}$ is the dipole size. In general, one is not able to
solve Eq.~(\ref{Bal1}) analytically. Only a ``piecewise'' expression
for $\alpha_s Y \gtrsim 1$ is known, and when translated to
$\varphi$ it reads
\begin{equation}\label{phiY}
    \varphi(k_,Y)=
    \begin{cases}
        \displaystyle{\frac{1}{\alpha_s} \ln
        \frac{Q_s^2}{k^2}}
        & \text{if\: $k \ll Q_s$}
        \vspace*{0.2cm}
        \\
        \displaystyle{\frac{1}{\alpha_s}
        \left(\frac{Q_s^2}{k^2} \right)^{\gamma_s}\!\!
        \left( \ln \frac{k^2}{Q_s^2} + \Delta \right)}
        & \text{if\: $k \gtrsim Q_s$}
        \vspace*{0.2cm}
        \\
        \displaystyle{\frac{Q_0^2}{k^2}\,
        {\rm I}_0\left(\sqrt{4 \abar Y \ln \frac{k^2}{Q_0^2}}\,
        \right)}
        & \text{if\: $k \gg Q_s$},
    \end{cases}
\end{equation}
where the dominant behavior of the saturation momentum is
$Q_s^2(Y)\!=\!\# Q_s^2(0) \exp [\abar \chi(\gamma_s)Y/\gamma_s]$,
with $\chi(\gamma)$ the eigenvalue of the BFKL equation and
$\gamma_s=0.628$ the associated anomalous dimension
\cite{GLR,IIM02,MT02,Tri03,MP0304b}. In Eq.~(\ref{phiY}), $\Delta$
is an undetermined constant and $\rm{I_0}$ is a modified Bessel
function of the first kind. From the first two pieces in this
equation, it is obvious that the solution exhibits geometrical
scaling \cite{GKS01,IIM02,MT02,Tri03,MP0304b,Lub01,RW04} below and
in a certain wide region above $Q_s$; it depends on $k$ and $Y$ only
through the combination $k^2/Q_s^2(Y)$. It is instructive to do a
first step in the non-linear evolution, valid so long as $Y \ll
1/\alpha_s$. To the order of accuracy and for momenta $k \lesssim
Q_s(A,Y)$, it is enough to evolve only the compact piece in
Eq.~(\ref{phi}) which will add a correction of order $Y$. This
correction contains power-law tails which are generated from the
tails of the evolution kernel. It is clear that, when $Y \gtrsim
1/\alpha_s$ all the components will be ``mixed'' and, unlike the
classical case, \emph{in the quantum case there is no compact
distribution for $k \lesssim Q_s(Y)$ and no parametric separation
between the solutions above and below $Q_s(Y)$}, as can be seen in
Eq.~(\ref{phiY}).

The analysis of the Cronin ratio is not trivial since we do not know
the solution in the whole $k$-$Y$ plane. Furthermore, given a
``point'' in this plane, the proton and the nucleus can be in
different phases, e.g.~the nucleus could be saturated while the
proton is still dilute. However, we can understand the generic
important features.

\noindent $\bullet$ The proton is ``less saturated'' than the
nucleus, since the initial proton scale $\sim \Lambda^2$ is much
smaller than the initial nuclear one $Q_s^2(A)$, and therefore the
available transverse space for the proton is larger. Thus, \emph{the
proton evolves faster than the nucleus and the ratio $\rpa$
decreases}. For example, along the particular line $k=Q_s(A,Y)$, one
has
\begin{equation}\label{RpAfixk}
    \frac{\dif \rpa}{\dif Y} < 0
    \quad \& \quad
    \rpa \xrightarrow[Y \rightarrow \infty]{}(\alpha_s A^{-1/3})^{1-\gamma_s}.
\end{equation}
\noindent$\bullet$ For fixed $Y$ and for extremely high momenta both
systems are dilute, described by the solution in the double
logarithmic approximation (the last piece in Eq.~(\ref{phiY})), and
the above mentioned difference in transverse space is now
unimportant. The ratio approaches 1 from \emph{below}, namely
\begin{equation}\label{RpafixY}
    \frac{\dif \rpa}{\dif k^2} \bigg|_{k^2 \gg Q_s^2(A,Y)} >0
    \quad \& \quad
    \rpa \xrightarrow[k^2 \rightarrow \infty]{}1^{-}.
\end{equation}
\noindent$\bullet$ The sum rule breaks down for any $Y>0$ because of
the correlations induced among the sources. The peak remains for a
while, since \emph{the sum rule is a sufficient but not a necessary
condition for the existence of a peak}.

\begin{figure}[t]
    \centerline{\resizebox{0.5\textwidth}{!}
    {\includegraphics{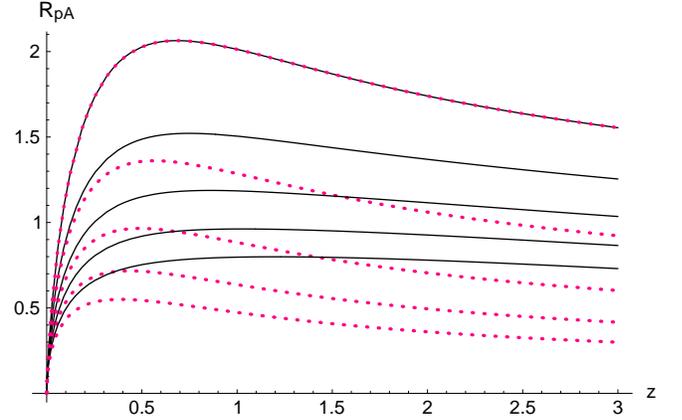}}}
    \caption{\sl
    From top to bottom, the Cronin ratio for $Y=0, 1/2,...,2$,
    below and near the saturation scale.
    Solid (black) lines correspond to an evolved nuclear wavefunction
    and dotted (red) to an unevolved one.}
    \label{fig4}
\end{figure}

\noindent $\bullet$ One can follow analytically the evolution of the
peak, as shown in Fig.~\ref{fig4}, until it becomes of order 1,
since this happens very fast due (again) to the large separation
between the scales $\Lambda^2$ and $Q_s^2(A)$. The nuclear
wavefunction is almost unevolved, while the proton is still dilute.
One finds
\begin{equation}\label{peak1}
    \rpa^{\rm max}=\mathcal{O}(1)
    \;\; {\rm when} \;\;
    Y=(1/4) \ln^2(1/\alpha_s) \ll 1/\alpha_s.
\end{equation}

\noindent$\bullet$ Even though smaller than 1, a peak persists under
further evolution until $Y \sim 1/\alpha_s$, when the power-law
tails will have ``washed-out'' the compact piece in the nuclear
wavefunction; \emph{the peak flattens out due to the nuclear
evolution} and the ratio becomes a monotonic function of $k^2$, that
is
\begin{equation}\label{nopeak}
    \frac{\dif \rpa}{\dif k^2}>0
    \quad {\rm when} \quad
    Y \gtrsim 1/\alpha_s.
\end{equation}
\noindent These features of saturation and the Cronin ratio
\cite{IIT04} extend previous discussions \cite{KLN03,KKT03}, agree
with the results obtained in numerical solutions \cite{AAKSW04}, and
remain qualitatively unaltered under a running coupling treatment
\cite{IIT04}.

\begin{figure*}[t]
    \centerline{\resizebox{1.0\textwidth}{!}
    {\includegraphics{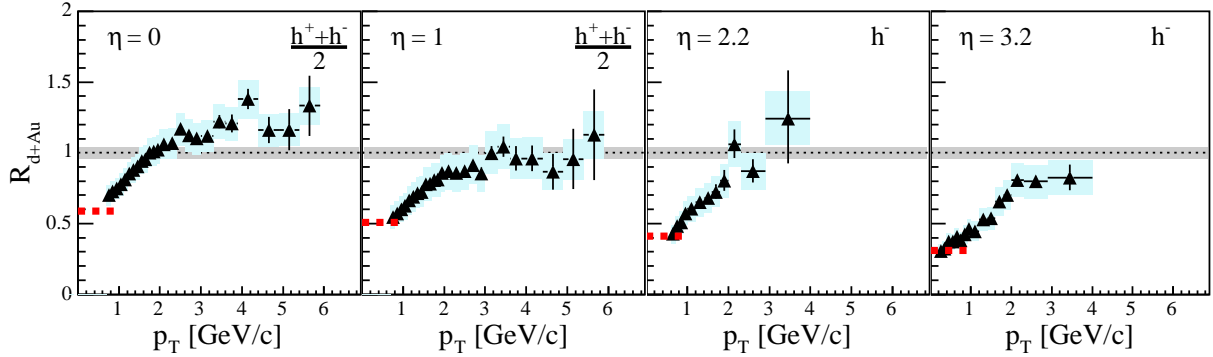}}}
    \caption{\sl The BRAHMS data \cite{BRAHMS04}:
    Nuclear modification factor for charged hadrons.}
    \label{fig5}
\end{figure*}

\section{Epilogue}
\label{epilogue} Saturation phenomena can play a significant role in
determining the produced particle spectra in high-energy heavy ion
collisions. In d-Au collisions at RHIC at BNL, final state
interactions are not important, and with $p_{\rm
\scriptscriptstyle{T}}$ and $\eta$ the transverse momentum and the
(pseudo)rapidity of the produced particle, one probes the nuclear
wavefunction at a value $x_{\rm Au} \simeq 2 |p_{\rm
\scriptscriptstyle{T}}| \exp[-\eta]/\sqrt{s_{\rm \scriptscriptstyle
{NN}}}$, where $\sqrt{s_{\rm \scriptscriptstyle {NN}}}$ is the
center of mass energy per nucleon. At current energies one expects
to reveal classical saturation properties in the mid-rapidity
region, while quantum saturation should be realized in the forward
one. Indeed, the CGC predictions (and postdictions) seem to be in
reasonable qualitative agreement with the RHIC data \cite{BRAHMS04}
shown in Fig.~\ref{fig5}. One should keep in mind that the gluon
occupation number and the particular ratio we studied are not
directly measurable quantities, and therefore any conclusion drawn
at the quantitative level might be misleading. Nevertheless, more
``refined'' quantities like, for example, the gluon production
\cite{KKT03,AAKSW04,BGV04ab} (and the corresponding ratio) and even
the charged hadron production \cite{KKT04}, are directly related to
the gluon occupation number and share the same features as those
presented in the previous sections. It could be very well the case
that the data for the nuclear modification factor shown in
Fig.~\ref{fig5} correspond to a manifestation of saturation in the
nuclear wavefunction.

\section*{Acknowledgements}

I wish to thank Edmond Iancu and Kazu Itakura, with whom the results
presented in this paper were obtained \cite{IIT04}, and Raju
Venugopalan for reading the manuscript.

\end{document}